\documentclass[a4paper]{article}%
\usepackage{amsmath}
\usepackage{mathrsfs}
\usepackage{array}
\usepackage{lipsum,appendix}
\usepackage{amsfonts}
\usepackage{amssymb}
\usepackage[lmargin=2.0cm,rmargin=3cm,tmargin=3.0cm,bmargin=2.0cm]{geometry} 
\usepackage{graphicx}%
\usepackage{textcomp}
\usepackage{gensymb}

\begin{document}

\title{An analytic solution to the spin flavor precession for solar Majorana neutrinos in the case of three neutrino generations} 
 
\author{Deniz Yilmaz\\ \textit{Department of Physics Engineering, Faculty of Engineering, Ankara University}\\ \textit{06100 Tandogan, Ankara, TURKEY}\\ \textit{e-mail:dyilmaz@eng.ankara.edu.tr} }

\maketitle

\begin{abstract}
The spin flavor precession (SFP) is investigated in the three neutrino generation case assuming that the neutrinos are Majorana type. 
Approximate analytical formulas including all transition magnetic moments are provided for the electron neutrino survival probability 
and $\nu_e \rightarrow \overline{\nu}_e$ transition probability in the SFP framework. 
The accuracy of the formulas is checked out for two different magnetic field profiles in the Sun. 
\end{abstract}

\section{INTRODUCTION}

The combined analysis of the solar neutrino experiments [1-8] and reactor antineutrino experiment [9-10] established to 
confirm the neutrino oscillation strongly indicates the so-called large mixing angle (LMA) region of the neutrino parameter space [11-16].
In a minimal extension of the standard model, neutrinos have mass; hence they also have magnetic moment.
In addition to the limits on the neutrino magnetic moments obtained by experimental and theoretical studies [17-23], an upper bound was recently obtained by 
GEMMA experiments: $\mu_{\nu}<2.9\times10^{-11}\mu_{B}$ at $90\%$ CL [24]. 
One can also find detailed analyses and discussions on neutrino magnetic moment in the literature [25-31]
When the neutrinos having nonzero magnetic moment propagate in a magnetic field, their spin can flip. 
Thus, a left handed neutrino becomes right handed neutrino, $\nu_{e_L}\rightarrow\nu_{e_R}$, which is deliberated as a possible solution to the solar neutrino deficit [32]. 
When the matter effect is included, then a left handed neutrino becomes right handed another type of neutrino: $\nu_{e_L}\rightarrow\nu_{\mu_R} \text{ or } \nu_{\tau_R}$ [33]. 
As distinct from the Dirac case in which right handed neutrino is considered as sterile which is not detectable by detectors,
in the Majorana case right handed neutrino is called antineutrino which can be detectable. 
This mechanism called as spin flavor precession (SFP) has been studied in different aspects [34-47].
In addition to the neutrino magnetic moment, a magnetic field profile in the Sun has to be choosen in order to carry out the SFP analysis quantitatively. 
The strength of the magnetic field in the Sun is limited by the standard solar model [48, 49] such as $\sim$ 20 G near the solar surface [50], 
20 kG - 300 kG at the convective zone [48] and $<  10^7 $ G at the solar center [48].
In this study two plausible profiles are considered as given in [51]; the first one is of Gaussian type having a peak at the bottom of the convective zone (Figure 1.a) and 
the second one is of the Wood-Saxon type being maximal at the centre of the Sun (Figure 1.b).   

In this paper, the SFP effect is studied in the case of three neutrino generations assuming that the neutrinos are Majorana type and 
the approximate analytical formula including all neutrino parameters and all types of neutrino magnetic moments is provided for the electron neutrino survival probability 
and $\nu_e \rightarrow \overline{\nu}_e$ transition probability.
The accuracy of the formula is checked out at the different $\theta_{12}$, $\delta m^2_{12}$ values and for two different magnetic field profiles in the Sun.
In section 2, the formalism of the SFP mechanism is examined for the three neutrino generations. 
The deduction of the approximate analytical formulas are given in the third and fourth sections. 
Results and Conclusion are presented in the last section.

\begin{figure}
[t]
\begin{center}
\includegraphics[height=6cm,width=9cm]{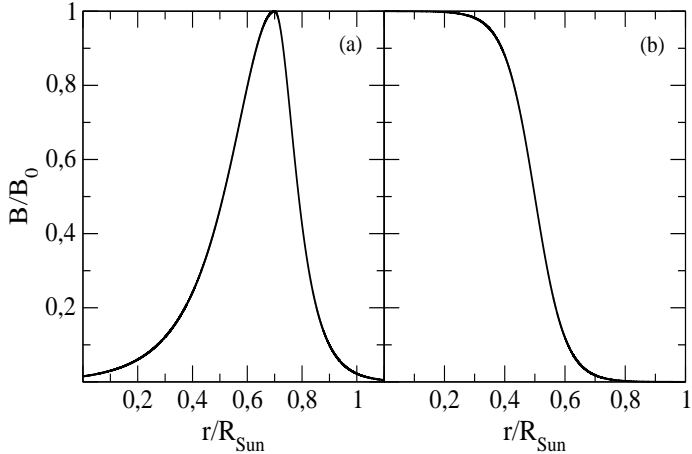}%
\caption{Magnetic field profiles: (a) Gaussian shape; (b) Wood-Saxon shape.} \label{fig:fig1}
\end{center}
\end{figure}

\section{Spin Flavor Precession For The Three Neutrino Generations}
In the case of three neutrino generations, the evolution equation for Majorana neutrinos passing through the matter and the magnetic field 
can be generated by using $6\times6$ rotational matrices consisting of the  $3\times3$ 
standard PMNS (Pontecorvo, Maki, Nakata, Sakata) mixing matrix [52]:   
\begin{equation}
T_{12}=\left(
\begin{array}
[c]{cc}%
R_{12} & 0  \\
0 & R_{12} \\
\end{array}
\right),\quad T_{13}=\left(
\begin{array}
[c]{cc}%
R_{13} & 0  \\
0 & R_{13} \\
\end{array}
\right),\quad T_{23}=\left(
\begin{array}
[c]{cc}%
R_{23} & 0  \\
0 & R_{23} \\
\end{array}
\right)
\end{equation}
here,
\begin{equation}
 R_{23}R_{13}R_{12}=\left(
\begin{array}
[c]{ccc}%
1 & 0 & 0  \\
0 &c_{23} & s_{23}   \\
0 & -s_{23} & c_{23}   \\
\end{array}\right)\left(
\begin{array}
[c]{ccc}%
c_{13} & 0 & s_{13}e^{-i\delta}  \\
0 & 1 & 0  \\
-s_{13}e^{i\delta} & 0 & c_{13} \\
\end{array}\right)\left(
\begin{array}
[c]{ccc}%
c_{12} & s_{12} & 0 \\
-s_{12} & c_{12}& 0  \\
0 & 0 & 1  \\
\end{array}\right) 
\end{equation}
where $c_{ij}=cos\theta_{ij}$ and $s_{ij}=sin\theta_{ij}$ and the $\delta$ is the CP-violating phase that we will ignore in our discussion.
Hereafter, we will use some useful abbreviations, such as:
\begin{equation}
 \begin{array}
[c]{c}%
 s^{2}_{ij}=sin^{2}\theta_{ij} \quad\quad c^{2}_{ij}=cos^{2}\theta_{ij}\\
 s2_{ij}=sin(2\theta_{ij}) \quad\quad c2_{ij}=cos(2\theta_{ij})\\ 
\Delta_{ij}=\dfrac{\delta m^{2}_{ij}}{2E}
\end{array}\quad \quad\quad \quad
\end{equation}
By taking $\Psi$ and $\overline{\Psi}$ as
\begin{equation}
\Psi= \left(
\begin{array}
[c]{c}%
\psi _{e}\\
\psi_{\mu}\\
\psi_{\tau}\\
\end{array}
\right),\quad
\overline{\Psi}= \left(
\begin{array}
[c]{c}%
\overline{\psi}_{e}\\
\overline{\psi}_{\mu}\\
\overline{\psi}_{\tau}\\
\end{array}
\right)
\end{equation}
the evolution equation for Majorana neutrinos passing through the matter and the magnetic field can be written as 
\begin{equation}
i\frac{d}{dt}\left(
\begin{array}
[c]{c}%
\Psi \\
\overline{\Psi}\\
\end{array}
\right) = \left[
T_{23}T_{13}T_{12} \left(
\begin{array}
[c]{cc}%
E  & 0  \\
0 & E  \\
\end{array}
\right)T^{\dagger}_{12}T^{\dagger}_{13}T^{\dagger}_{23}
+\left(
\begin{array}
[c]{cc}%
V  & 0  \\
0 & -V  \\
\end{array}
\right)+\left(
\begin{array}
[c]{cc}%
0 & BM  \\
-BM& 0 \\
\end{array}
\right) \right]\left(
\begin{array}
[c]{c}%
\Psi \\
\overline{\Psi}\\
\end{array}
\right)
\end{equation}
where $E $, $V$ and $M$ are the $3\times3$ submatrices:
\begin{equation}
E=\left(
\begin{array}
[c]{ccc}%
E_{1} & 0 & 0  \\
0 & E_{2} & 0  \\
0 & 0 & E_{3}  \\
\end{array}
\right),\text{ }
V=\left(
\begin{array}
[c]{ccc}%
V_{c}+V_{n} & 0 & 0  \\
0 & V_{n} & 0  \\
0 & 0 & V_{n}  \\
\end{array}
\right),\text{ }
M=\left( \begin{array}
[c]{ccc}%
0 & \mu_{e\mu} & \mu_{e\tau} \\
-\mu_{e\mu} & 0 & \mu_{\mu\tau} \\
-\mu_{e\tau} & -\mu_{\mu\tau} & 0 \\
\end{array}\right)
\end{equation}
here $\mu_{i j} $ is the transition magnetic moment ( $i$ and $j$ denote the $e$, $\mu$, $\tau$).
The evolution equation we end up with is
\begin{equation}
i\dfrac{d}{dt}\left(
\begin{array}
[c]{c}%
\varphi\\
\overline{\varphi}\\
\end{array}
\right)= \mathscr{H} \left(
\begin{array}
[c]{c}%
\varphi\\
\overline{\varphi}\\
\end{array}
\right)
\end{equation}
here,
\begin{equation}
\varphi=\left(
\begin{array}
[c]{c}%
\varphi_{e}\\
\varphi_{\mu}\\
\varphi_{\tau}\\
\end{array}
\right)=\left(
\begin{array}
[c]{c}%
 c_{13} \psi _e-s_{13} \tilde{\psi }_{\tau } \\
 \tilde{\psi }_{\mu } \\
 s_{13} \psi _e+c_{13} \tilde{\psi }_{\tau } \\
\end{array}
\right)
\end{equation}
and
\begin{equation}
 \begin{array}
[c]{c}%
 \tilde{\psi}_{\mu}=c_{23} \psi_{\mu}-s_{23} \psi_{\tau}\\
\tilde{\psi}_{\tau}=s_{23} \psi_{\mu}+c_{23} \psi_{\tau}\\
\end{array}
\end{equation}
Identical expressions are true for antineutrinos (just put a bar above).
In the evalution equation above, $\mathscr{H}$ is 
\begin{equation}
\mathscr{H}=\left(
\begin{array}{cc}
 H & B M^\prime  \\
-BM^\prime  &\overline{H} \\
\end{array}
\right)
\end{equation}
here , after we have substracted off overall phase from the Hamiltonian, $H$, $\overline{H}$ and $M^\prime$ are

\begin{equation}
H=\left(
\begin{array}{ccc}
\frac{1}{2} \left(c_{13}^2 V_{c}-c2_{12} \Delta_{21}\right) & \frac{s2_{12} \Delta_{21}}{2} & c_{13} s_{13} V_{c} \\
\frac{s2_{12} \Delta_{21}}{2} & \frac{1}{2} \left(-c_{13}^2 V_{c}+c2_{12} \Delta_{21}\right) & 0\\
 c_{13} s_{13} V_{c} & 0 & V_{c}-\frac{3 c_{13}^2 V_{c}}{2}+\frac{1}{2} \left(\Delta_{31}+\Delta_{32}\right) 
\end{array}
\right)
\end{equation}

\begin{equation}
\overline{H}=\left(
\begin{array}{ccc}
 D_{11} & \frac{1}{2}s2_{12} \Delta_{21} & -c_{13} s_{13} V_{c}  \\
 \frac{1}{2}s2_{12} \Delta_{21} & D_{22} & 0\\
 -c_{13} s_{13} V_{c} & 0 & D_{33} 
\end{array}
\right)
\end{equation}
together with the diagonal elements
\begin{eqnarray}
\begin{array}{l}
D_{11}=  -\frac{3}{2} c_{13}^2 V_{c}-2 V_{n}-\frac{1}{2}c2_{12}\Delta_{21}   \\
D_{22}=  -\frac{1}{2}c_{13}^2 V_{c}-2 V_{n}+\frac{1}{2}c2_{12}\Delta_{21} \\
D_{33}=  \frac{1}{2}\left(-2+c_{13}^2\right)V_{c}-2 V_{n}+\frac{1}{2}\left(\Delta_{31}+\Delta_{32}\right) \\
\end{array}
\end{eqnarray}
and
\begin{equation}
M^\prime=\left(
\begin{array}{ccc}
 0 & \mu_{12} & \mu_{13} \\
- \mu_{12} & 0 & \mu_{23}\\
- \mu_{13} & - \mu_{23}  & 0
\end{array}
\right)
\end{equation}
here we defined
\begin{eqnarray}
\begin{array}{l}
\mu_{12}= c _{13} c_{23} \mu_{e \mu}-c_{13} s_{23}  \mu_{e \tau}+s_{13}  \mu_{\mu \tau} \\
\mu_{13}= s_{23}  \mu_{e \mu}+c_{23}  \mu_{e \tau}\\
\mu_{23}= -c_{23}  s_{13}  \mu_{e \mu}+s_{13}  s_{23}  \mu_{e \tau}+c_{13}  \mu_{\mu \tau}. \\
\end{array}
\end{eqnarray}
The matter potentials for the charged and the neutral current are given by
\begin{equation}
\begin{array}{c}
 V_c=\sqrt{2}G_F N_e\\
 V_n=-\frac{G_F}{\sqrt{2}} N_n\\
\end{array}
\end{equation}
here $G_F$ is Fermi constant, $N_e$ and $N_n$ are electron and neutron density, respectively.

\section{Deduction of The Electron Neutrino Survival Probability}

We start with the evolution equation obtained at the section 2 for Majorana neutrinos passing through the matter and the magnetic field: 

\begin{equation}
i\dfrac{d}{dt}\left(
\begin{array}
[c]{c}%
\varphi\\
\overline{\varphi}\\
\end{array}
\right)= \mathscr{H} \left(
\begin{array}
[c]{c}%
\varphi\\
\overline{\varphi}\\
\end{array}
\right)
\end{equation}
here $\varphi$ and $\overline{\varphi}$ denote the three neutrino and anti-neutrino flavor part, respectively. 
The matrix of $\mathscr{H}$ can be splitted into the two parts as a matter and a magnetic part:
\begin{equation}
\mathscr{H}=\mathscr{H}_{M}+\mathscr{H}_{B}
\end{equation}
here $\mathscr{H}_{M}$ and $\mathscr{H}_{B}$ are
\begin{equation}
 \mathscr{H}_{M}=\left[
\begin{array}{cc}
 H & 0\\
 0 &  \overline{H}  \\
\end{array}
\right],
\end{equation}
\begin{equation}
 \mathscr{H}_{B}=\left[
\begin{array}{cc}
 0 &  B M^\prime \\
 - B M^\prime & 0  \\
\end{array}
\right].
\end{equation}
We are going to solve the three-by-three blocks and use these solutions to solve the six-by-six matrix later. 

Since the upper diagonal part of $\mathscr{H}_{M}$ is related to the neutrinos, the evolution equation for the neutrinos is
\begin{equation}
i\dfrac{d}{dt}\left(
\begin{array}
[c]{c}%
\varphi_{e}\\
\varphi_{\mu}\\
\varphi_{\tau}\\
\end{array}
\right)=H \left(
\begin{array}
[c]{c}%
\varphi_{e}\\
\varphi_{\mu}\\
\varphi_{\tau}\\
\end{array}
\right).
\end{equation}
Let' s split $H$ into two parts as well:
\begin{equation}
 H=H^{0}+H^{1}
\end{equation}
here $H^{0}$ is
\begin{equation}
H^{0}= \left(\begin{array}{ccc}
\frac{1}{2} \left(c_{13}^2 V_{c}-c2_{12} \Delta_{21}\right) & \frac{s2_{12} \Delta_{21}}{2} & 0 \\
 \frac{s2_{12} \Delta_{21}}{2} & \frac{1}{2} \left(-c_{13}^2 V_{c}+c2_{12} \Delta_{21}\right) & 0 \\
 0 & 0 & b
\end{array}
\right)
\end{equation}
where $b$ is 
\begin{equation}
b=V_{c}-\frac{3 c_{13}^2 V_{c}}{2}+\frac{1}{2} \left(\Delta_{31}+\Delta_{32}\right).
\end{equation}
The evolution operator $U_{H}$ for $H$ satisfies the equation 
\begin{equation}
i\frac{d}{dt} U_{H}=HU_{H}.
\end{equation}
The solution to this equation can be sought by taking $U_{H}=U^{0}_{H}U^{1}_{H}$.
Since the equation, $i\frac{d}{dt} U^{0}_{H}=H^{0}U^{0}_{H}$, associated with $H^{0}$ is the standard 
$2\times2$ MSW equation with an independent third flavor, the solution can be chosen as
\begin{equation}
U^{0}_{H}=\left(
\begin{array}{ccc}
 \psi _1& -\psi^{*}_{2}& 0 \\
 \psi _2  & \psi^{*}_{1} & 0 \\
 0 & 0 & \beta \\
\end{array}
\right)
\end{equation}
where $\psi _1 (t)$, $\psi _2 (t)$ are $2\times2$ MSW solutions with the initial conditions $\psi _1 (t=0) =1$, $\psi _2 (t=0) =0$ [11] and $\beta$ is
\begin{equation}
 \beta =e^{-i\int b \, dt}.
\end{equation}
We now can find the complete solution to all of $H$ by looking at $H^1$:
\begin{equation}
i \frac{\partial U^{1}_{H}}{\partial t}={U^{0}_{H}}^{\dagger } H^{1} U^{0}_{H} U^{1}_{H}=h(t)U^{1}_{H}
\end{equation}
Because $\theta_{13}$ is small, we can apply an approximation to the solution of this equation:
\begin{equation}
U^{1}_{H}=exp (-i\int _0^t h(t^\prime)dt^\prime)= \left[1-i\int _0^t h(t^\prime)dt^\prime-\frac{1}{2}\int _0^t \int_0^{t^\prime} h(t^\prime)h(t^{\prime \prime}) dt^\prime dt^{\prime \prime}+\text{...}\right]
\end{equation}
Then we have the whole solution for $H$
\begin{equation}
 U_{H}=U^{0}_{H}U^{1}_{H}.
\end{equation}

The antineutrino part of $\mathscr{H}_{M}$, $\overline{H}$, can be similarly solved.
After $\overline{U}_{\overline{H}}$ is obtained, the solution matrix for the $\mathscr{H}_{M}$ can be written as:
\begin{equation}
U_{M}=\left[
\begin{array}{cc}
U_{H} & 0 \\
0 & \overline{U}_{\overline{H}} \\
\end{array}
\right]
\end{equation}
Since the total evolution is characterized by $U=U_{M}U_{B}$, we need $U_{B}$ which is the solution matrix of $\mathscr{H}_{B}$. By using the equation satisfied by the evolution operator of $\mathscr{H}$
\begin{equation}
i\frac{d }{d t}U=\mathscr{H}U
\end{equation}
we can get
\begin{equation}
i\dfrac{d}{d t} U_{B}=(U_{M}^{\dagger }\mathscr{H}_{B}U_{M}) U_{B}=h_{b}(t)U_{B}
\end{equation}
Because $\mu B$ is small, $U_{B}$ can also be found by applying an approximation up to the second order in $\mu B$
\begin{equation}
U_{B}=\left[1-i\int_0^t h_{b}(t^\prime) dt^\prime-\frac{1}{2}\int _0^t \int_0^{t^\prime} h_{b}(t^\prime)h_{b}(t^{\prime \prime}) dt^\prime dt^{\prime \prime}+\text{...}\right]
\end{equation}
The state of the system evolves with a unitary operator from the initial state
\begin{equation}
\left(
\begin{array}
[c]{c}%
\varphi (t)\\
\overline{\varphi} (t)\\
\end{array}
\right)= U \left(
\begin{array}
[c]{c}%
\varphi(t=0)\\
\overline{\varphi}(t=0)\\
\end{array}
\right)
\end{equation}
with 
\begin{equation}
U=U_{M}U_{B}=\left(
\begin{array}{cc}
 \text{A} & \text{C} \\
 \text{D} & \text{B}\\
\end{array}
\right)
\end{equation}
where $A$, $B$, $C$ and $D$ are $3 \times 3 $ matrices. The electron neutrino amplitude can be written from the equation (8) as 
\begin{equation}
\psi_{e}=c_{13}\varphi_e+s_{13}\varphi_{\tau},
\end{equation}
Since the elements of $\overline{\varphi} (t)$ at $t=0$ is zero, it is enough to look at the A matrix only to obtained $\varphi_e$ and $\varphi_{\tau}$.
\begin{equation}
\left(
\begin{array}
[c]{c}%
\varphi_{e}\\
\varphi_{\mu}\\
\varphi_{\tau}\\
\end{array}
\right)=\left(
\begin{array}{ccc}
 A_{11} & A_{12} & A_{13}   \\
 A_{21} & A_{22} & A_{23}   \\
 A_{31} & A_{32} & A_{33}   \\
\end{array}
\right)\left(
\begin{array}{c}
 c_{13}\\
 0\\
 s_{13}
\end{array}
\right)
\end{equation}
Therefore, one can find
\begin{equation}
 \begin{array}
[c]{c}%
 \varphi _{e}=A_{11}c_{13}+A_{13}s_{13} \\
 \varphi _{\tau}=A_{31}c_{13}+A_{33}s_{13}.
\end{array}\quad \quad\quad \quad
\end{equation}

The highly oscillating integrations coming out in the solution matrix elements are ignored. 
However, stationary phase approximation method [53] can be used for the other integrals in which the SFP resonance width is considerably small as mentioned in [42].
After the terms that have higher order than $s^2_{13}$ are ignored, only $A_{11}$, $A_{31}$ and $A_{33}$ matrix elements left: 
\begin{eqnarray}
\begin{array}{l}
A_{11}=\psi_{1}(t) \left(1 - \frac{1}{4} \frac{2\pi \Gamma^2_{\mu B} }{\vert d(\chi-2\kappa)/dt \vert_{(\chi-2\kappa)=0}} \vert\psi_{1}(t_R)\vert ^2 \right) \\ 
A_{31}=-ic_{13} s_{13} e^{-i\int_{0}^t b \, dt^\prime}\int_{0}^t dt^\prime V_c (t^\prime) e^{i\int_{0}^{t^\prime} b \, dt^{\prime \prime}} \psi_{1}(t^\prime)  \\
A_{33}=e^{-i\int_{0}^t b \, dt^\prime}
\end{array}
\end{eqnarray}
here, 
\begin{eqnarray}
\begin{array}{l}
\Gamma_{\mu B}= \mu_{eff}B\\
\mu_{eff}=c _{13} c_{23} \mu_{e \mu}-c_{13} s_{23} \mu_{e \tau}+s_{13} \mu_{\mu \tau} \\
\kappa=\frac{\Delta_{21}}{2} c2_{12}\\
\chi=\frac{G_f}{\sqrt{2}} \left(2N_e-2N_n\right)\\
\end{array}
\end{eqnarray}

Substituting $\varphi _{e}$ and $\varphi _{\tau}$ and the terms, $A_{11}$, $A_{31}$ and $A_{33}$ into $\psi_e$ we obtain   
\begin{equation}
\begin{array}{l}
\psi_e=\psi_{1}(t) c_{13}^2 \left(1 -\frac{1}{4}\frac{2\pi \Gamma^2_{\mu B} }{\vert d(\chi-2\kappa)/dt \vert_{(\chi-2\kappa)=0}} \vert\psi_{1}(t_R)\vert ^2 \right) \\
\quad\quad-i c^2_{13} s^2_{13}e^{-i\int_{0}^t b \, dt^\prime}\int_{0}^t dt^\prime V_c (t^\prime) e^{i\int_{0}^{t^\prime} b \, dt^{\prime \prime}} \psi_{1}(t^\prime)+s_{13}^2e^{-i\int_{0}^t b \, dt^\prime}
\end{array}
\label{eq:xdef}
\end{equation}
here, integral can be solved by using the same method given in [11].

One can finally get the electron survival probability for three neutrino generations in the SFP framework by ignoring the terms that have higher order than $(\mu B)^2$ 
\begin{equation}
\begin{array}{l}
 P_{3\times3}(\nu_e \rightarrow \nu_e, \mu B\neq0)= c^4_{13} P_{2\times2}(\nu_e \rightarrow \nu_e \text{ } \text{ with }\text{ }  N_e c^2_{13}, \mu B=0) \left(1-\frac{1}{2}\frac{2\pi \Gamma^2_{\mu B} }{\vert d(\chi-2\kappa)/dt \vert_{(\chi-2\kappa)=0} } \vert\psi_{1}(t_R)\vert ^2  \right)\\
\quad\quad \quad\quad\quad\quad+s^4_{13}\left[1+ 2\xi c^2_{13}+\xi^2 c^4_{13}\right]
\end{array}
\end{equation}
where,
\begin{equation}
\xi=\frac{V_c(t=0)}{\Delta_{31}}.
\end{equation}

\section{Analytical Expression for $\nu_e \rightarrow \overline{\nu}_e$ Transition Probability}
If the neutrinos are assumed to be Majorana type, $\nu_e$ changes to $\overline{\nu}_{\mu,\tau}$ inside the Sun by SFP. 
After Sun, $\overline{\nu}_{\mu,\tau}$ transforms to $\overline{\nu}_e$ via vacuum oscillation:
\begin{align*}
\nu_e  \stackrel{SFP}{\rightarrow}  \overline{\nu}_{\mu,\tau} \stackrel{V_{osc}}{\rightarrow} \overline{\nu}_e. 
\end{align*}
Hence, the electron antineutrino flux, $\Phi_{\overline{\nu}_{e}}(E) $, on Earth is given by 
\begin{equation}
\Phi_{\overline{\nu}_{e}}(E) =\Phi_{\nu_{e}}(E)\times P(\nu_e \rightarrow \overline{\nu}_e)
\end{equation}
where $\Phi_{{\nu}_{e}}(E) $ is the solar electron neutrino flux with energy E.
Therefore, one needs the $\nu_e \rightarrow \overline{\nu}_e$ transition probability to find the electron antineutrino flux on Earth:
\begin{equation}
P(\nu_e \rightarrow \overline{\nu}_e)= P(\nu_e \rightarrow \overline{\nu}_{\mu,\tau};SFP) \times P(\overline{\nu}_{\mu,\tau} \rightarrow \overline{\nu}_e;VacuumOsc.)
\end{equation}
here, $P(\overline{\nu}_{\mu,\tau} \rightarrow \overline{\nu}_e;VacuumOsc.)$ is the well known vacuum oscillation probability given as 
\begin{equation}
P(\overline{\nu}_{\mu,\tau} \rightarrow \overline{\nu}_e;VacuumOsc.)=sin^{2}\theta_{12}sin^2(\frac{\delta m^2_{12}}{4E}R)\stackrel{averaging}{\rightarrow}\frac{1}{2}sin^{2}\theta_{12}. 
\end{equation}
and $P(\nu_e \rightarrow \overline{\nu}_{\mu,\tau};SFP)$ is the $\nu_e \rightarrow \overline{\nu}_{\mu,\tau}$ transition probability:
\begin{equation}
P(\nu_e \rightarrow \overline{\nu}_{\mu,\tau};SFP)=  \vert\ \overline{\psi}_{\mu,\tau} \vert ^2 
\end{equation}
here, $\overline{\psi}_{\mu,\tau}$ can be found with the solution of antineutrino part in equation (19) by using the same method given in section 3:
\begin{equation}
\overline{\psi}_{\mu}=-i c_{13}  \left( \frac{2\pi \Gamma^2_{\mu B} }{\vert d(\chi-2\kappa)/dt \vert_{(\chi-2\kappa)=0}} \right)^{1/2} \psi_{1}(t_R) \overline{\psi}_{2}(t_R) \left(c_{23}\overline{\psi}^{*}_{1}(t)+s_{23}s_{13}\overline{\psi}^{*}_{2}(t) \right)
\end{equation}
and
\begin{equation}
\overline{\psi}_{\tau} =\overline{\psi}_{\mu} (c_{23}\rightarrow-s_{23},\text{ } s_{23}\rightarrow c_{23}).
\end{equation}

\section{Results and Conclusions}

\begin{figure}
[t]
\begin{center}
\includegraphics[height=14cm,width=14cm]{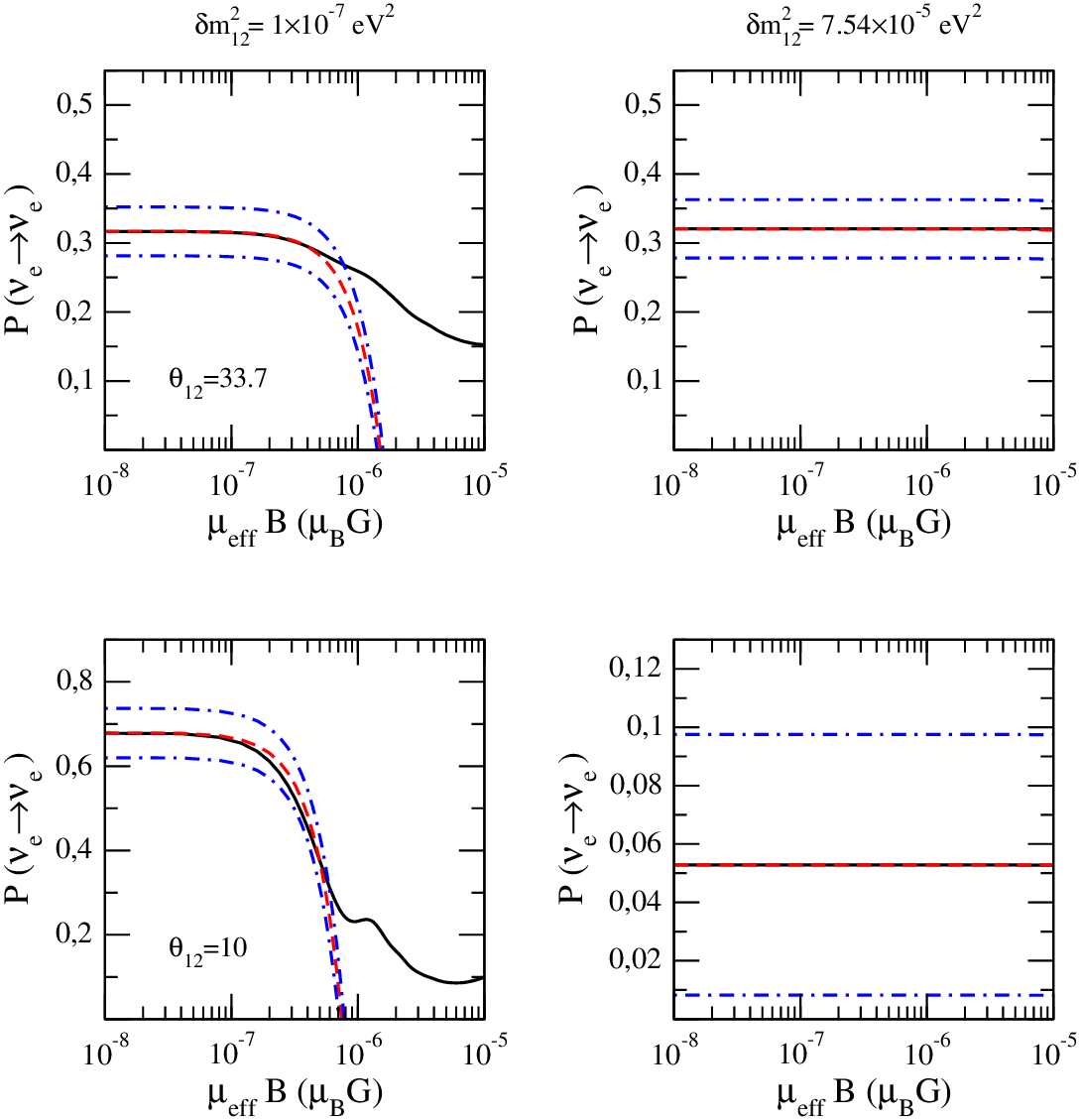}%
\caption{Survival probabilities for the 10 MeV neutrino energy at different $\theta_{12}$ and $\delta m^2_{12}$ values for Gaussian shape of magnetic field profile.
While the solid lines show the results obtained numerically, the dashed lines show the result obtained from the approximate analytical expression.
The dotted-dashed lines show the errors. Each column (row) uses the same  $\delta m^2_{12}$ ($\theta_{12}$) values. } \label{fig:fig2}
\end{center}
\end{figure}

\begin{figure}
[t]
\begin{center}
\includegraphics[height=14cm,width=14cm]{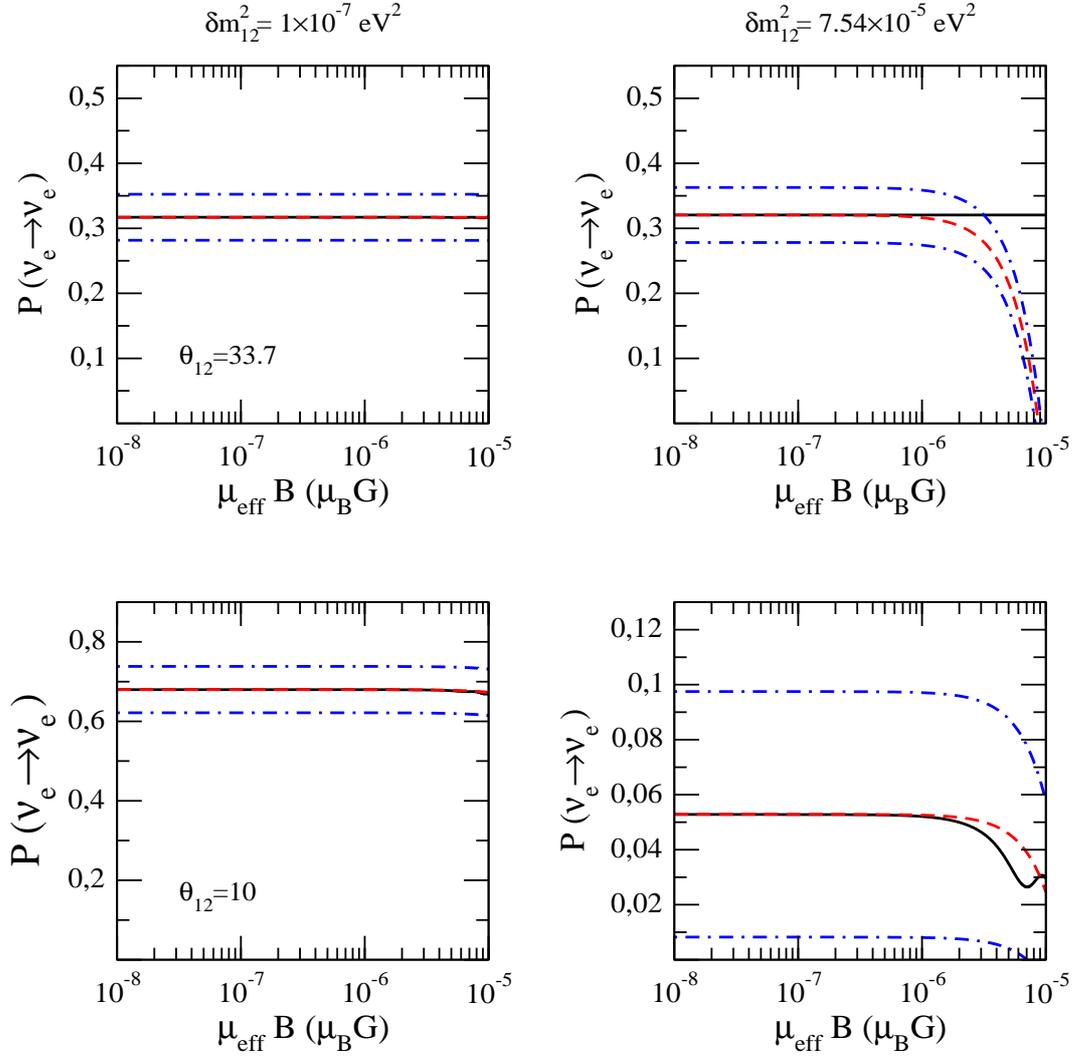}%
\caption{Survival probabilities for the 10 MeV neutrino energy at different $\theta_{12}$ and $\delta m^2_{12}$ values for Wood-Saxon shape of magnetic field profile.
While the solid lines show the results obtained numerically, the dashed lines show the result obtained from the approximate analytical expression.
The dotted-dashed lines show the errors. Each column (row) uses the same  $\delta m^2_{12}$ ($\theta_{12}$) values.} \label{fig:fig3}
\end{center}
\end{figure}

\begin{figure}
[t]
\begin{center}
\includegraphics[height=12cm,width=14cm]{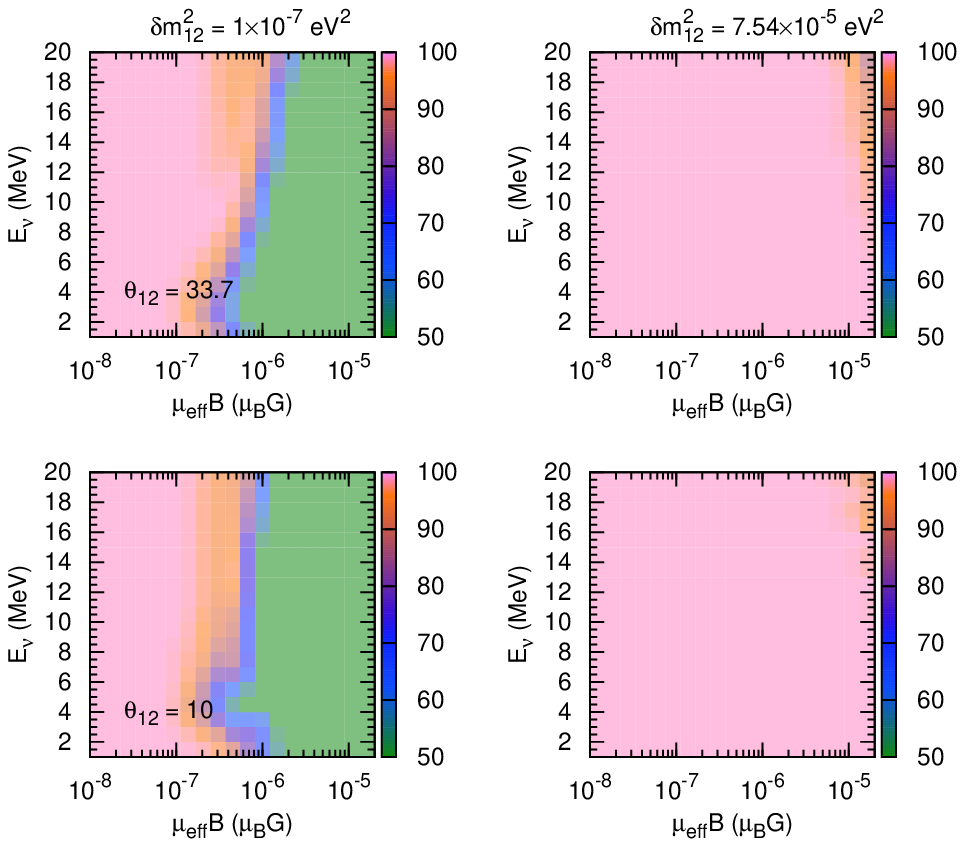}%
\caption{Percent accuracy regions at different $\theta_{12}$ and $\delta m^2_{12}$ values 
for Gaussian shape of magnetic field profile. The accuracy rates are given between $100\%$ and $50\%$.
Each column (row) uses the same  $\delta m^2_{12}$ ($\theta_{12}$) values.} \label{fig:fig4}
\end{center}
\end{figure}

\begin{figure}
[t]
\begin{center}
\includegraphics[height=12cm,width=14cm]{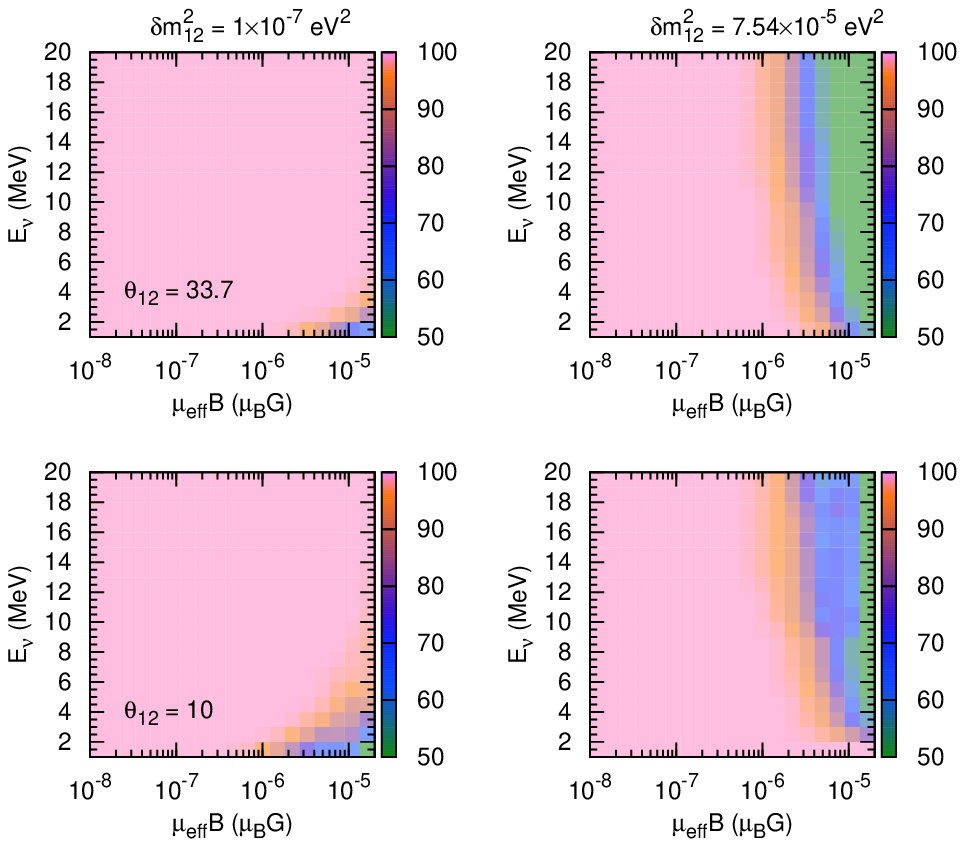}%
\caption{Percent accuracy regions at different $\theta_{12}$ and $\delta m^2_{12}$ values
for Wood-Saxon shape of magnetic field profile. The accuracy rates are given between $100\%$ and $50\%$.
Each column (row) uses the same  $\delta m^2_{12}$ ($\theta_{12}$) values.} \label{fig:fig5}
\end{center}
\end{figure}

\begin{figure}
[t]
\begin{center}
\includegraphics[height=6cm,width=14cm]{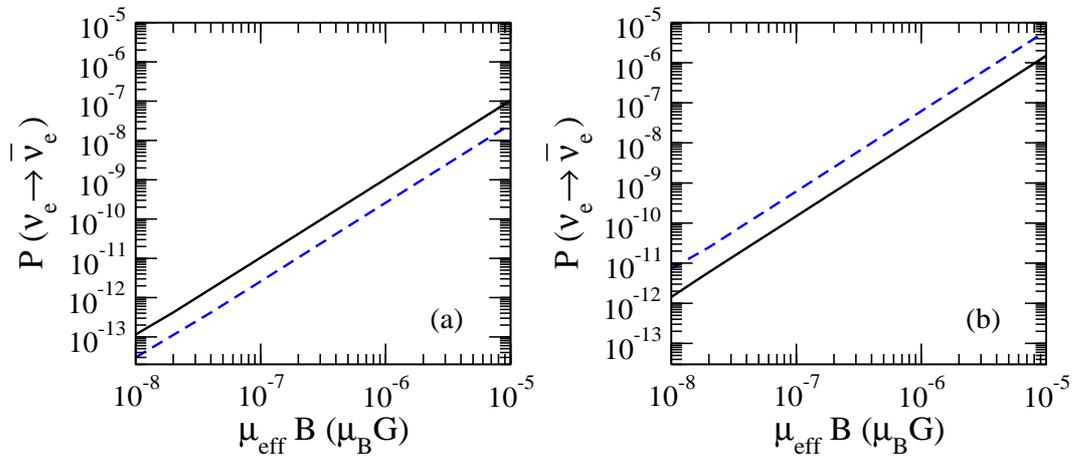}%
\caption{$\nu_e \rightarrow \overline{\nu}_e$ transition probabilities for 2 MeV neutrino energy and at best fit LMA values of $\theta_{12}$ and $\delta m^2_{12}$ for 
Gaussian shape (a) and Wood-Saxon shape (b) of magnetic field profiles.
While the solid lines show the results obtained numerically, the dashed lines show the result obtained from the approximate analytical expression.} \label{fig:fig6}
\end{center}
\end{figure}

In this paper, we have examined the SFP mechanism in the three neutrino generations and obtained 
approximate analytical formulas including all neutrino parameters and all types of neutrino magnetic moments. 
The accuracy of the approximate solution obtained by the formulas is checked out by comparing with the exact solution obtained numerically by 
diagonalizing the Hamiltonian in (7) for two different magnetic field profiles in the Sun.
In the calculations, Gaussian (Figure 1.a) and the Wood-Saxon shape (Figure 1.b) of magnetic field profiles extending over the entire Sun are choosen [51].
The figures (Figure 2 - Figure 6) are plotted as a function of the product $\mu_{eff}B$, since $\mu_{eff}$ and $B$ appeared together in the survival probability
expression in equation (43). $\mu_{eff}$ given explicitly in equation (41) includes three transition magnetic moments 
with the upper bound $\mu_{ij} \lesssim 10^{-11}\mu_{B}$ ($i$ and $j$ denote $e$, $\mu$, $\tau$) [31]. 
Results are presented at different $\theta_{12}$, $\delta m^2_{12}$ values. The best fit values of all neutrino parameters and their errors are taken from [52].  

Electron neutrino survival probabilities obtained by using exact (solid lines) and approximate solution (dashed lines) for the 10 MeV neutrino energy are shown 
in Figure 2 and Figure 3 with the errors (dotted-dashed lines) for Gaussian and the Wood-Saxon shape of magnetic field profiles, respectively. 
In these figures one can see at what values of $\mu_{eff}B$ the approximate solution works well.  

Figure 4 and Figure 5 show the percent accuracy regions at different $\theta_{12}$ and $\delta m^2_{12}$ values for 
Gaussian and the Wood-Saxon shape of magnetic field profiles, respectively. In these figures, compared to the results from the exact solution, 
how reliable the approximate analytical formula is shown. 
The accuracy rates are given between $100\%$ and $50\%$. The results obtained from the formula are quite compatible with the exact solution results 
(almost in $99.9$ percent accuracy) at $\delta m^2_{12}=7.54\times 10^{-5} eV^2$ for Gaussian magnetic field profile and at $\delta m^2_{12}=1\times 10^{-7} eV^2$ for 
Wood-Saxon magnetic field profile for almost all $\mu_{eff}B$ values and neutrino energies. 
Besides, for the MSW-LMA best fit values (upper right panels of each figures), we have $99.9$ percent accuracy as well nearly up to the 
$1-2\times 10^{-6}\mu_B G$ value of $\mu_{eff}B$ which is a sufficiently high value for the Sun 
at all neutrino energies for both magnetic field profile. 
Additionally, it is seen that for $\delta m^2_{12}=7.54\times 10^{-5} eV^2$, the formula is also highly reliable at low neutrino energies even for high enough values of $\mu_{eff}B$.

In Figure 6, $\nu_e \rightarrow \overline{\nu}_e$ transition probabilities are shown for 2 MeV neutrino energy and at best fit LMA values of $\theta_{12}$ and $\delta m^2_{12}$ for 
Gaussian shape (a) and Wood-Saxon shape (b) of magnetic field profiles.
It can be seen that the results obtained from the analytical expression (dashed lines) are compatible with the ones obtained numerically (solid lines) 
for both magnetic field profiles.
This may allow the expression to be used in the calculations of the solar antineutrino flux on Earth. 

In conclusion, even though the evolution equation can be solved numerically, one might need to have an approximated analytical solution to see the behaviour of the probabilities 
in SFP framework without performing detailed numerical analysis. The dependence of the analytical probability expressions 
on the neutrino parameters can be seen from the expressions between the equation (43) and equation (50).
Moreover, the formulas derived here can also be useful when the data obtained by new 
solar neutrino experiments is analyzed in the SFP framework.  

\section*{Acknowledgement}
I would like to thank A. Baha Balantekin and Annelise Malkus for the insightful discussions and their invaluable suggestions.
I also thank Y\"{O}K (Council of Higher Education) for the grant to support my research at the University of Wisconsin, Madison.

\end{document}